\documentclass[aps,numerical,superscriptaddress,floatfix,reprint,groupedaddres,longbibliography]{revtex4-2}

\usepackage{graphicx,color}
\usepackage{amsmath,amssymb,amsfonts,bm}
\usepackage{mathbbol}
\usepackage{amsfonts}
\usepackage{mathrsfs}
\usepackage[colorlinks=true,linkcolor=blue,plainpages=false]{hyperref}
 \usepackage{relsize}
 
 \usepackage{amssymb}
\usepackage{mathbbol}
\usepackage{amsfonts}
\usepackage{mathrsfs}
\usepackage{epsfig,bm,dcolumn}
\usepackage{graphicx}
\usepackage{color}
\usepackage{amsmath}
\usepackage{dcolumn}
\usepackage{overpic}
\usepackage{hyperref}
\usepackage{slashed}
\usepackage{hyperref}

\newcommand{\be}{\begin{equation}}
\newcommand{\ee}{\end{equation}}
\newcommand{\ba}{\begin{eqnarray}}
\newcommand{\ea}{\end{eqnarray}}
\newcommand{\ban}{\begin{eqnarray*}}
\newcommand{\ean}{\end{eqnarray*}}

\def\v2{\mbox{$v_2$}}

\begin{document}

\title{Accessing Topological Fluctuations of Gauge Fields  with Chiral Magnetic Effect }
\medskip

\author{Anping Huang}
\affiliation{School of Nuclear Science and Technology, University of Chinese Academy of Sciences, Beijing 100049, China.}

\author{Shuzhe Shi}
\affiliation{Department of Physics, McGill University, 3600 University Street, Montreal, QC, H3A 2T8, Canada.}

\author{Shu Lin}
\affiliation{School of Physics and Astronomy,
Sun Yat-Sen University, Zhuhai, 519082, China.
}

\author{Xingyu Guo}
\affiliation{Guangdong Provincial Key Laboratory of Nuclear Science, Institute of Quantum Matter, 
South China Normal University, Guangzhou 510006, China.} 
\affiliation{Guangdong-Hong Kong Joint Laboratory of Quantum Matter, South China Normal University, Guangzhou 510006, China.}

 \author{Jinfeng Liao}
\email{liaoji@indiana.edu}
\affiliation{ Physics Department and Center for Exploration of Energy and Matter,
Indiana University, 2401 N Milo B. Sampson Lane, Bloomington, IN 47408, USA.} 

 \date{\today}


\begin{abstract}
Gauge fields provide the fundamental interactions in the Standard Model of particle physics. Gauge field configurations with nontrivial topological windings are known to play crucial roles in many important phenomena, from matter-anti-matter asymmetry of today's universe to spontaneous chiral symmetry breaking in strong interaction. Their presence is however elusive for direct detection in experiments. Here we show that measurements of the chiral magnetic effect (CME) in heavy ion collisions can be used for accessing the topological fluctuations of the non-Abelian gauge fields in the Quantum Chromodynamics (QCD). To achieve this, we implemented a key ingredient, the stochastic dynamics of gauge field topological fluctuations, into a state-of-the-art framework for simulating the CME in these collisions. This new framework provides the necessary tool to quantify initial topological fluctuations from any definitive CME signal to be extracted from experimental data. It also reveals  a universal scaling relation between initial topological fluctuations and particle multiplicity produced in the corresponding collision events. 
\end{abstract}

\maketitle

\section*{Introduction}


The fundamental structures and interactions of all visible matter in our universe are well described by the Standard Model of particle physics together with gravitation.  In the Standard Model, gauge fields arising from underlying gauge symmetry principles provide the strong, weak and electromagnetic forces that hold the physical world together. A fascinating aspect of gauge fields is related to gauge field configurations with nontrivial topological windings. They  emerge in various physical systems of different dimensions and play crucial roles in many important phenomena. Well-known examples include magnetic vortices in superconductors, monopoles in the electroweak theory, as well as instantons and sphalerons in non-Abelian gauge theory.  See reviews in e.g.\cite{tHooft:1999cgx,Schafer:1996wv,Shuryak:2021vnj,Shifman:2012zz}.  

Let us focus on the instantons and sphalerons of non-Abelian gauge theories in four-dimensional spacetime~\cite{Belavin:1975fg,Klinkhamer:1984di}. These topological configurations  constitute crucial  ingredients for our very existence. On the very small scale of subatomic dynamics, 
they lead to the spontaneous chiral symmetry breaking in the strong interaction vacuum~\cite{Schafer:1996wv,Shuryak:2021vnj,Greensite:2011zz,Diakonov:2009jq} and thus 
provide the dominant source of mass for visible matter. On the very large scale of cosmic evolution, they allow the violation of baryon number conservation and thus possible emergence of the large matter-anti-matter asymmetry in today's universe~\cite{Rubakov:1996vz,Arnold:1987mh}.  The essence of such topological configurations is the tunneling transitions across energy barriers between the topologically distinct vacuum sectors of a non-Abelian gauge theory characterized by different Chern-Simons numbers. In doing so, they themselves  ``twist'' topologically  around spacetime boundary and can be characterized by their {\em topological winding numbers}, defined as:  
\begin{eqnarray}  \label{eq_qw}
Q_w = \int d^4 x \,  q(x) = \int d^4 x \left[- \frac{g^2\epsilon^{\mu\nu\rho\sigma}}{32\pi^2}\text{Tr} \left\{G_{\mu\nu}G_{\rho\sigma} \right\}\right ] \,\,\, 
\end{eqnarray}
where the integrand $q(x)$ is the local topological charge density of a given gauge field configuration described by gauge field strength tensor $G_{\mu\nu} (x)$ with $g$ being the corresponding coupling constant.  

Despite their significance,   the topological configurations are elusive experimentally. A direct detection of their presence and consequences in laboratories  could substantially advance our understanding of the underlying tunneling mechanism that is at the heart of quark confinement and baryon asymmetry, but has not been achieved so far. A concrete proposal~\cite{Kharzeev:1998kz,Kharzeev:2001ev,Kharzeev:2004ey,Voloshin:2004vk} toward this goal is to look for the parity-odd ``bubbles'' (i.e. local domains) arising from the topological transitions of gluon fields in Quantum Chromodynamics (QCD). Specifically, these bubbles could occur in the hot quark-gluon plasma (QGP) created by relativistic heavy ion collisions. The parity-odd nature of such a bubble can be quantified by the macroscopic {\em chirality} $N_5$ generated for the light quarks in the plasma. Indeed, this is enforced by the famous chiral anomaly relation for  massless fermions (i.e. the light flavor quarks in the case of QCD): 
\begin{eqnarray}  \label{eq_anomaly_local} 
 \partial_\mu J^\mu_5 &=& 2q(x) =- \frac{g^2}{16\pi^2} \epsilon^{\mu\nu\rho\sigma} \text{Tr}\left\{G_{\mu\nu}G_{\rho\sigma} \right\} \,\, , \\
N_5 &\equiv& N_R - N_L = 2 Q_w \label{eq_anomaly_global}   \,\, .
\end{eqnarray}
In the above $J^\mu_5$ is the local chiral or axial current for each quark flavor while the Eq.(\ref{eq_anomaly_global}) is the spacetime-integrated version of Eq.(\ref{eq_anomaly_local}), with $N_R$ and $N_L$ being the number of right-handed (RH) and left-handed (LH) quarks. This latter equation has its deep mathematical root in the celebrated Atiyah-Singer index theorem and physically means that each topological winding generates two units of net chirality per flavor of light quarks. Therefore, measuring the net chirality of QGP provides a unique way of directly accessing the fluctuations of gluon field  topological windings  in heavy ion collision experiments. 
 {Note though the winding number $Q_w$ in principle can not be an accurately measurable observable according to the Wigner-Araki theorem, for which the practical constraint could nevertheless be made arbitrarily small~\cite{WA2011}. What one can hope for is to quantify the average winding number that occurs in the fireball spacetime volume from event-by-event topological fluctuations.}

The quark net chirality, however, is also challenging to detect due to the fact that the QGP born from collisions would expand, cool down and eventually transition into a low temperature hadron phase where the spontaneous breaking of chiral symmetry makes the net chirality unobservable. Fortunately, there is a way out by virtue of the so-called {\em chiral magnetic effect (CME)}~\cite{Kharzeev:2007jp,Fukushima:2008xe}. The CME is an anomalous transport phenomenon where an electric current $\mathbf{J}$ is induced along an external magnetic field $\mathbf{B}$ under the presence of net chirality in a system with massless fermions of charge $Q_e$:  
\begin{eqnarray}  \label{eq_cme} 
  \mathbf{J} &=& \frac{Q_e^2}{2\pi^2} \mu_5  \mathbf{B}  \, .
\end{eqnarray}
The $\mu_5$ is a chiral chemical potential that quantifies the net chirality $N_5$. The study of CME has attracted significant interests and activities from  a broad range of  physics disciplines such as high energy physics, condensed matter physics, astrophysics, cold atomic gases, etc. See recent reviews in e.g.  \cite{Kharzeev:2020jxw,Armitage:2017cjs,Burkov:2017rgl,Miransky:2015ava,Fukushima:2018grm}. In the context of heavy ion collisions, the CME current (\ref{eq_cme}) leads to a charge separation in the quark-gluon plasma that results in a specific hadron emission pattern and can be measured via charge-dependent azimuthal correlations~\cite{Voloshin:2004vk}.    Extensive experimental efforts have been carried out over the past decade to look for its signatures at the Relativistic Heavy Ion Collider (RHIC) and the Large Hadron Collider (LHC): see more details in recent reviews~\cite{Kharzeev:2015znc,Zhao:2018skm,Li:2020dwr}.

In short, there is a promising pathway for experimental probe of gauge field topology in heavy ion collisions: the winding number $Q_w$ of gluon fields $\Rightarrow$ net chirality of quarks 
$\Rightarrow$ CME current $\Rightarrow$ correlation observables. 
Here we demonstrate, for the first time, how this strategy actually works quantitatively for counting the topological winding numbers of gauge field with  measurements of the CME. 
 {
To be precise, the $Q_w$ fluctuates from event to event with equal chance of being positive or negative and what one can hope for is to determine its average variance $\mathlarger{\mathlarger{\mathlarger{\sigma}}}_w \equiv \sqrt{\langle Q_w^2\rangle_{event}}$ associated with such fluctuations over the fireball spacetime domain.  This quantity $\mathlarger{\mathlarger{\mathlarger{\sigma}}}_w$ 
is ultimately related to the two-point functions of the topological charge density and is what one aims to extract in this study.}
This has become possible due to two developments. Experimentally, great progress has been made in data analysis methods to separate background contamination and extract the CME signal with much reduced uncertainty~\cite{Zhao:2018skm,Li:2020dwr}. Theoretically, we have developed a new  framework for accurately  computing the CME transport  while taking into account the  stochastic dynamics of gluon field topological fluctuations during the  evolution in these collisions.

\begin{figure}[!hbt]
	\begin{center}
		\includegraphics[width=3.3in]{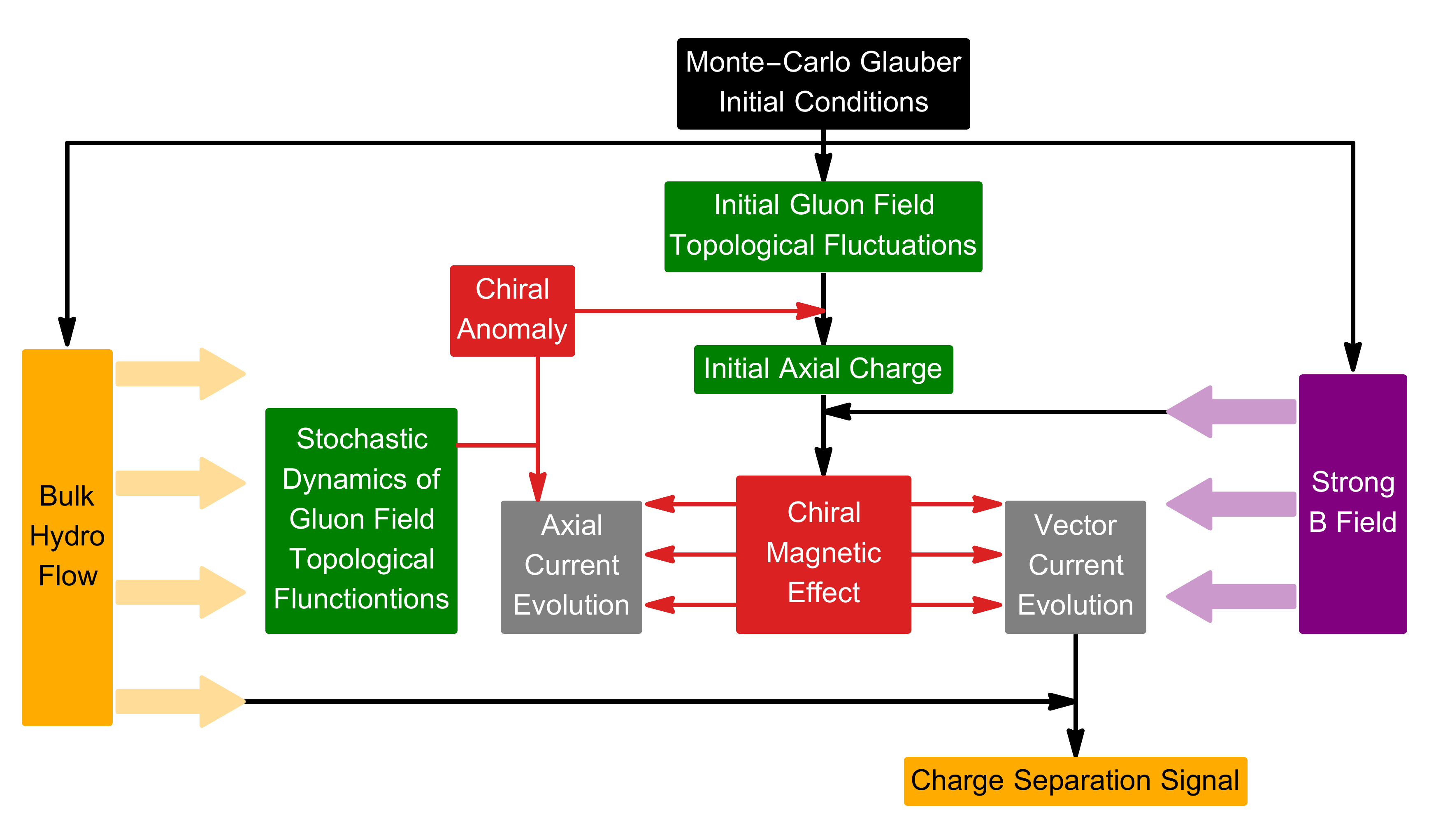}
		\caption{An illustration of our framework: anomalous-viscous fluid dynamics (AVFD) with stochastic dynamics of gauge field topological fluctuations. See Method section for details.  }
		\vspace{-0.5cm}
		\label{fig_1}
	\end{center}
\end{figure}

The various ingredients and flow chart  of our framework is illustrated in Fig.~\ref{fig_1}. This is built upon a state-of-the-art modeling tool, the anomalous-viscous fluid dynamics (AVFD)~\cite{Shi:2019wzi,Shi:2017cpu,Jiang:2016wve}, which provides the quantitative link by simulating the CME transport during the dynamical evolution of a heavy ion collision from initial topological winding $Q_w$ to final experimental signal. However, random topological fluctuations occur both at the very beginning of a collision and during the course of its evolution.  They result in flipping of chirality and impose important impact on the CME transport, which was previously neglected. To precisely count the initial topological windings, one must account for such missing key ingredient. In this work we've successfully implemented  both the event-by-event initial topological fluctuations and the stochastic dynamics of gluon field topological fluctuations into the AVFD simulations (as indicated by the green blocks in Fig.~\ref{fig_1}), thus paving the way for counting $Q_w$ with CME measurements. See Method section for a more detailed discussion about this framework.

\section*{Method}


Here we present details of the various components (as illustrated in Fig.~\ref{fig_1}) for the computational   framework developed for this study.  

Heavy ion collisions at RHIC and LHC energies are well described by relativistic viscous hydrodynamic simulations, which have been thoroughly vetted with extensive experimental data. The ``backbone'' of our framework for describing such bulk evolution is based on the MUSIC (2+1)D code package~\cite{Schenke:2010nt,Schenke:2010rr}  with initial time $\tau_0=0.4\rm fm/c$ and shear viscosity parameter $\eta/s =0.08$. The event-wise initial conditions (e.g. for entropy density profiles) of the bulk hydro are generated with Monte-Carlo Glauber simulations. 

The anomalous-viscous fluid dynamics (AVFD)~\cite{Shi:2017cpu,Jiang:2016wve,Shi:2019wzi} is the key component for   implementing the dynamical CME transport  in the realistic environment of a relativistically expanding viscous fluid. This state-of-the-art tool numerically solves the   anomalous hydrodynamic equations for the coupled evolution of quarks' vector and axial currents together   on top of the bulk collective flow. Starting from a given axial charge initial condition,   the magnetic-field-induced CME currents eventually lead to a charge separation effect in the fireball, which turns into a dipole term $\pm a_1 \sin (\phi-\Psi_{RP})$ in the azimuthal angle distribution of positive/negative charged hadrons with respect to the reaction plane orientation $\Psi_{RP}$.  
Such a dipole signal is experimentally measurable through a difference between same-sign (SS) and opposite-sign (OS) charged hadron pair correlations, $H^{SS-OS} = \langle 2 a_1^2\rangle $.   
In short, the AVFD is a hydrodynamic realization of CME for quantifying its signal in heavy ion collisions. For more details, see refs.~\cite{Shi:2017cpu,Jiang:2016wve,Shi:2019wzi}. It may be noted that the lifetime of the magnetic field is a major source of theoretical uncertainty. To nail this down would require quantitative understanding of dynamically evolving magnetic field in the medium along the line of studies in e.g. \cite{Gursoy:2018yai,Yan:2021zjc}. The influence of this factor on CME signal is thoroughly investigated in  \cite{Shi:2017cpu}. We follow   \cite{Shi:2017cpu} to use a lifetime of $0.6\ \rm fm$ which is reasonable and supported by phenomenological analysis~\cite{Guo:2019joy}.  
 
 For the goal of this work, one crucial new ingredient has been introduced into our framework. During the hydrodynamic evolution, there exist random topological fluctuations of the gluon fields that would necessarily influence the axial current evolution. These fluctuations eventually amount to a relaxation effect toward equilibrium with vanishing topological charge on long time scale. To account for such effect, one needs to introduce the resulting relaxation term into the anomalous hydrodynamic  equation for the  axial current~\cite{Iatrakis:2014dka,Iatrakis:2015fma,Lin:2018nxj,Liang:2020sgr}:   
 \begin{eqnarray} \label{eq_nq}
\partial_{\mu} J^{\mu}_{f, 5} &=& - \frac{N_{c}Q^{2}_f}{2\pi^{2}}E\cdot B - \frac{n_{f,5}}{\tau_{cs}} \, . 
\end{eqnarray} 
In the above, 
$J^{\mu}_{f, 5}$ is the axial current of each quark flavor with electric charge $Q_f$ and color number $N_c=3$, while $n_{f,5}$ is the corresponding axial charge density.  The $E\cdot B$ in Eq.(\ref{eq_nq}) comes from the Abelian anomaly due to electromagnetic fields. The stochastic dynamics of gluon topological charge density $q(x)$ gives rise to a new contribution i.e. the second term in Eq.(\ref{eq_nq}), where the $\tau_{cs}$ is an important relaxation time   for the topological charge fluctuations. Its physical meaning is simple: over this time scale, the $q$ approaches equilibrium value. The $\tau_{cs}$ is controlled by the   Chern-Simons diffusion rate $\Gamma_{cs}$, i.e. 
\begin{eqnarray} 
\tau_{cs} = \frac{\chi\,T}{2N^{2}_{f}\Gamma_{cs}} \, . 
\end{eqnarray}
In the above $\chi$ is the total quark number susceptibility, $T$ is the temperature and $N_f=3$ is the light flavor number.  

Some discussions are now in order concerning the two key quantities here, i.e. $\tau_{cs}$ and $\Gamma_{cs}$. 
First, it is useful to compare $\tau_{cs}$ with different time scales involved in the problem. The microscopic scale relevant for bulk equilibrium in heavy ion collisions is presumably on the order of $\tau_{micro}\sim 0.1 \rm fm/c$ while the magnetic field lifetime scale is on the order of $\tau_{B}\sim 1 \rm fm/c$. In the scenario of $\tau_{cs}\ll \tau_{micro}$, the topological fluctuations behave as random noise that would wash out any meaningful CME transport on macroscopic scale. On the other hand, if $\tau_{cs} \gg \tau_B$ then the relaxation of initial topological fluctuations would be too slow to have an impact on the CME transport.  It is only when the $\tau_{cs}$ is on a par with $\tau_{B}$ that the stochastic dynamics of topological charges will play an important role. Therefore, a good estimate of $\tau_{cs}$ is crucial, which however is highly nontrivial and strongly dependent on the estimate of $\Gamma_{cs}$. 
For the latter, the best theoretical guidance is the perturbative result from \cite{Moore:2010jd}: $\Gamma_{cs} \approx 30(\alpha_{s}T)^{4}$. One has to recognize that the applicability of this result, when extrapolated to the regime relevant for heavy ion collisions, is just anyone's guess. It is therefore to clearly demonstrate the significant theoretical uncertainty present in this parameter and its consequences for the outcome of our calculations. In Table.~\ref{table_1}, we list the values of $\Gamma_{cs}$ and $\tau_{cs}$ for a wide range of choices for $\alpha_s$ based on the aforementioned formula. Needless to say, this formula may become totally inapplicable when $\alpha_s$ increases beyond certain range. However, we notice that the resulting $\tau_{cs}$ value (which is the actual input into our simulation framework)  already spans a rather generous range of possibilities. Comparing simulation results from these different choices will help calibrate the theoretical uncertainties associated with the relaxation time scale of topological charges, as we shall discuss later in the Results section.  While such a widespread range of $\tau_{cs}$ appears superficially discouraging, it shall nevertheless be noted that many studies of the initial stages in heavy ion collisions have found that the $\alpha_s=0.3$ is quite reasonable. Since the CME transport occurs mainly during the early stage, it is therefore a plausible choice to use this value for estimating $\tau_{cs}$ in the present work. 
Furthermore, the large uncertainty helps remind us that the very initial motivation of studying CME is to utilize this phenomenon for providing unique experimental constraints on such a highly uncertain property of gauge fields. 

\vspace{0.1in}
	\begin{table}[!hbt] 
	\begin{center}
		\begin{tabular}{| c  | c | c |  }
			\hline     $\alpha_s$ value   &  $\Gamma_{cs}$ (in GeV$^4$)  &     $\tau_{cs}$ (in fm/c)      \\
			\hline    0.1    &  $2.43\times 10^{-5}$    &     113      \\
			\hline    0.3    &   $1.97\times 10^{-3}$    &     1.40      \\			
			\hline    0.5    &  $1.52\times 10^{-2}$    &     0.18     \\
			\hline
		\end{tabular}
		\caption{The values of two key parameters $\Gamma_{cs}$ (in GeV$^4$) and  $\tau_{cs}$ (in fm/c)   at temperature  $T=300\rm MeV$ for different choices of input $\alpha_s$ value.}\vspace{-0.15in}
		\label{table_1}
\end{center}
	\end{table}

It may also be noted that the small but finite masses for the light flavor quarks (i.e. u, d, and s quarks) would also contribute to the diffusion rate. However recent analysis~\cite{Hou:2017szz} has shown convincingly that such mass contribution, even for the strange quarks with $\sim 100\rm MeV$ mass, is smaller than the above Chern-Simons diffusion rate by a few orders of magnitude and thus negligible. That means the strange quarks will also contribute to the anomalous transport just like the u and d flavors. 

Last but not least, it is important to properly generate the initial conditions for the axial charge density based on the initial topological charge density of the gluon fields. At the very early stage of high energy heavy ion collisions, many   flux tubes of  strong chromo electric and magnetic fields in parallel or anti-parallel configurations are formed. In such a glasma picture~\cite{Lappi:2006fp,Gelis:2010nm,Fukushima:2011nq,Gelis:2012ri}, these flux tubes extend along the collision beam axis and are localized on the transverse plane. The chromo field strength inside the tubes is on the order of $Q_s^2$ while their transverse size is on the order of $1/Q_s$, with $Q_s$ being the saturation scale.  Depending on whether the chromo electric and magnetic fields are in parallel or anti-parallel configurations, each flux tube possesses randomly positive or negative topological charge density. They seed the generation of initial axial charge density $n_{5,i}$ in the collisions~\cite{Kharzeev:2001ev,Hirono:2014oda,Lappi:2017skr}. We develop the following procedure to sample the initial axial charge density. For each collision event, we randomly sample a total of $N_{coll}$ glasma flux tubes on the transverse plane, where $N_{coll}$ is the binary collision number for each event generated from Monte Carlo Glauber simulations. For the $i$-th tube located at position $\mathbf{x}_i$, the chromo fields generate the following local axial charge upon integrating the Eq.(\ref{eq_anomaly_local}) up to the hydro initial proper time $\tau_0$: 
\begin{eqnarray}
n_{5,i}(\tau_0,x,y)=(\pm) \cdot\lambda\frac{Q_s^{2}}{8\pi^{2}}\tau_0 \, \left [\frac{1}{2\sigma^{2}}e^{-\frac{(\mathbf{x}-\mathbf{x}_{i})^{2}}{2\sigma^{2}}} \right ] 
\end{eqnarray}  
where $x,y$ are transverse spatial coordinates.  
In the above, the plus or minus sign is randomly sampled for each tube. 
 For our computations we use an average value of $Q_s^2= 1.5\ \rm GeV^2$  in the above for collisions at RHIC energy~\cite{Lappi:2006fp,Gelis:2010nm,Fukushima:2011nq,Gelis:2012ri}. The Gaussian  factor in the bracket represents a smearing of the generated axial charge over a spread size of $\sigma$. In our calculations we vary $\sigma$ in a reasonable range between $(0.5\sim 1)\rm fm$ as an estimator of systematic uncertainty due to the initial condition.  The $\lambda$ is a dimensionless strength parameter reflecting the fact that we only know these chromo fields up to the order of magnitude. Clearly $\lambda$ controls the amount of initial topological windings and consequentially the axial charges. As a last step, we superpose all the flux tubes in a given event together to obtain the overall axial charge initial profile: $n_{5}(\tau_0) = \sum_{i=1}^{N_{coll}} n_{5,i}$. This can then be used as the initial condition for solving previously discussed evolution equation for the axial current in our framework. 
Finally one can integrate the $n_5$ over the fireball to obtain the total axial charge $N_5$ in each event. 

Procedurally, one varies the value of control parameter $\lambda$ and perform event-by-event simulations to generate the CME signal. This establishes a mapping between the initial topological fluctuations $\mathlarger{\mathlarger{\mathlarger{\sigma}}}_w$ and the final CME signal   $H^{SS-OS}$. Then by comparison with experimentally extracted value for $H^{SS-OS}$, one can determines the corresponding $\mathlarger{\mathlarger{\mathlarger{\sigma}}}_w$. It may be worth mentioning that the correlation measurements typically contain both CME signal and background contamination. Methods have been developed and demonstrated to be able to extract the signal part~\cite{Bzdak:2012ia,Xu:2017qfs,Voloshin:2018qsm,Abdallah:2021itw,Lacey:2020arl,Tang:2019pbl,Choudhury:2021jwd,Christakoglou:2021nhe}.


\section*{Results}
 %

\begin{figure}[!hbt]
	\begin{center}
		\includegraphics[width=2.7in]{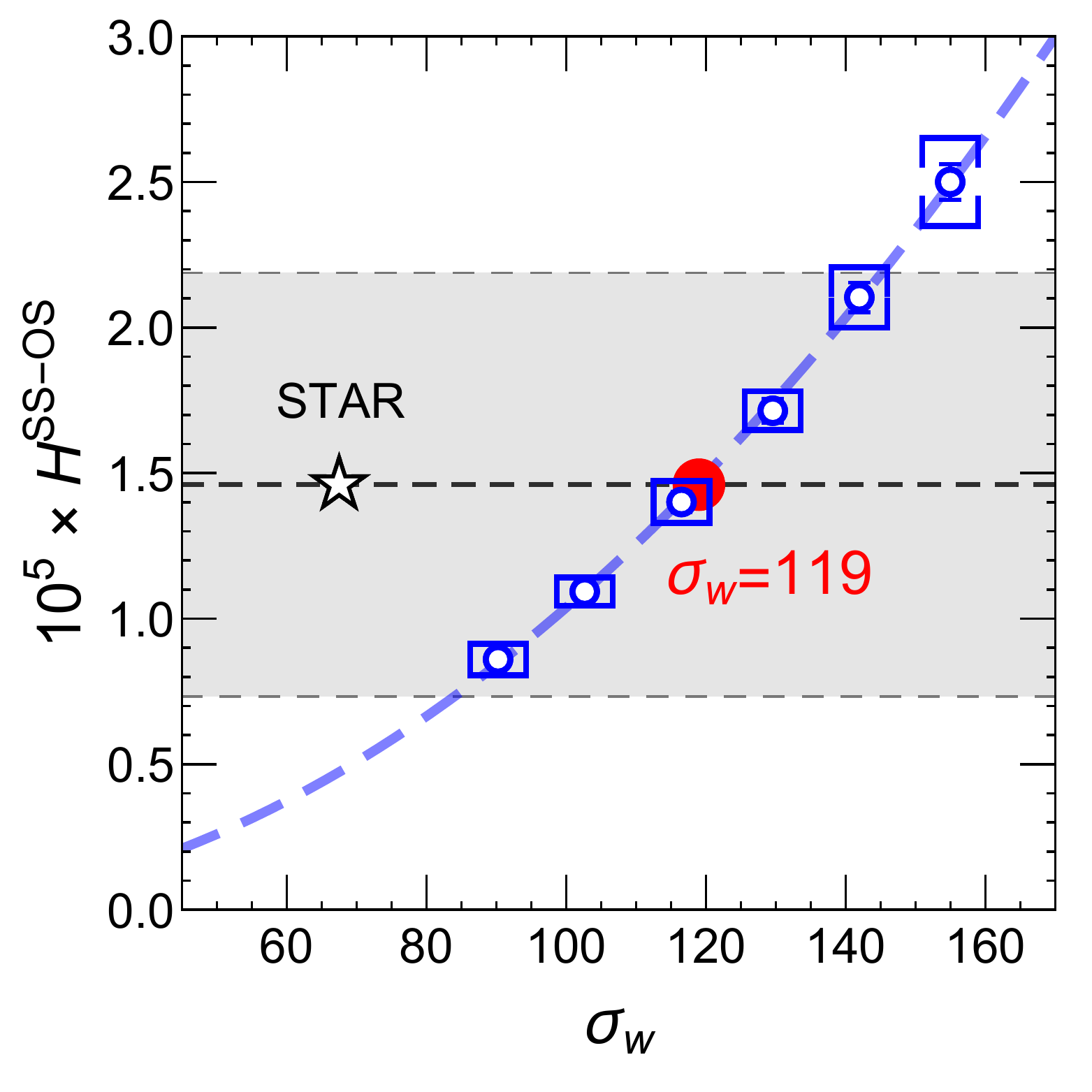}
		\caption{The  dependence of CME-induced correlation signal $H^{SS-OS}$ on the initial topological fluctuations 
		$\mathlarger{\mathlarger{\mathlarger{\sigma}}}_w$.   The blue circles are from numerical simulations, along with statistical error bars and systematic error boxes.   The dashed blue curve shows a quadratic fitting with $H^{SS-OS}= 1.038\times 10^{-9}\, {\mathlarger{\mathlarger{\mathlarger{\sigma}}}_w}^2$.  The star symbol and grey band show the current experimental constraints from analysis of the STAR collaboration data. The red dot at $\mathlarger{\mathlarger{\mathlarger{\sigma}}}_w=119$ indicates the value of 
		${\mathlarger{\mathlarger{\mathlarger{\sigma}}}_w}$  that gives a correlation signal in consistence with experimental data. 
		}
		\vspace{-0.5cm}
		\label{fig_2}
	\end{center}
\end{figure}

Using the above framework, we first show results for the AuAu collisions at $200$GeV beam energy and  $(20\sim50)\%$  centrality in Fig.~\ref{fig_2}, where the dependence of CME-induced correlation signal $H^{SS-OS}$ on the initial topological fluctuations $\mathlarger{\mathlarger{\mathlarger{\sigma}}}_w$ is shown. (For this calculation the $\tau_{cs}$ is calculated with 
$\alpha_s=0.3$.) The blue circles (along with statistical uncertainty bars and systematic uncertainty boxes) are obtained from numerical simulations by varying the control parameter $\lambda$ for initial chirality generation (--- see Method section for details). A theoretically-expected quadratic dependence is clearly observed, with the dashed blue curve showing an excellent fitting: $H^{SS-OS}= 1.038\times 10^{-9}\, {\mathlarger{\mathlarger{\mathlarger{\sigma}}}_w}^2$.  The star symbol and grey band show the current experimental constraints from analysis of the STAR collaboration data~\cite{Abelev:2009ac,Zhao:2018blc,Zhao:2018skm,Li:2020dwr}. Comparison between calculations and measurements suggest an optimal value of $\mathlarger{\mathlarger{\mathlarger{\sigma}}}_w=119$ (as indicated by the red dot). This demonstrates how the initial topological fluctuations can be quantitative extracted from any definitive CME signal from future experimental measurements. 

To demonstrate the impact of the stochastic dynamics, we have compared the current result with a reference calculation in which the stochastic dynamics is turned off and the initial axial charges do not dissipate away. We find that the correlation signal $H^{SS-OS}$ is reduced by a factor of 3.6 as compared with the reference calculation. Clearly the impact of stochastic topological fluctuations is significant and must be accounted for. To calibrated the uncertainty associated with the $\tau_{cs}$ estimate, we've also done calculations with different input $\alpha$ values. We find that $H^{SS-OS}= 3.578\times 10^{-9}\, {\mathlarger{\mathlarger{\mathlarger{\sigma}}}_w}^2$ for $\alpha_s=0.1$ and $H^{SS-OS}= 7.4 \times 10^{-11}\, {\mathlarger{\mathlarger{\mathlarger{\sigma}}}_w}^2$ for $\alpha_s=0.5$. The comparison confirms the sensitivity of the calculated signal to the choice of $\tau_{cs}$,  which also implies that future measurement of CME signal could help constrain this key property of gauge field topological fluctuations. 

As noted before, the finite quark masses have only negligible contributions to the axial charge dynamics and the strange quarks are expected to participate in CME transport just as the u and d flavors. A nontrivial consequence of this fact is a large charge correlation signal for strange mesons such as charged kaons, predicting a ratio  $H^{SS-OS}_{K^\pm} / H^{SS-OS}_{\pi^\pm} \approx 1.2$ which can be tested with CME measurements for identified hadrons. 

 \begin{figure}[!hbt]
	\begin{center}
			\includegraphics[width=2.7in]{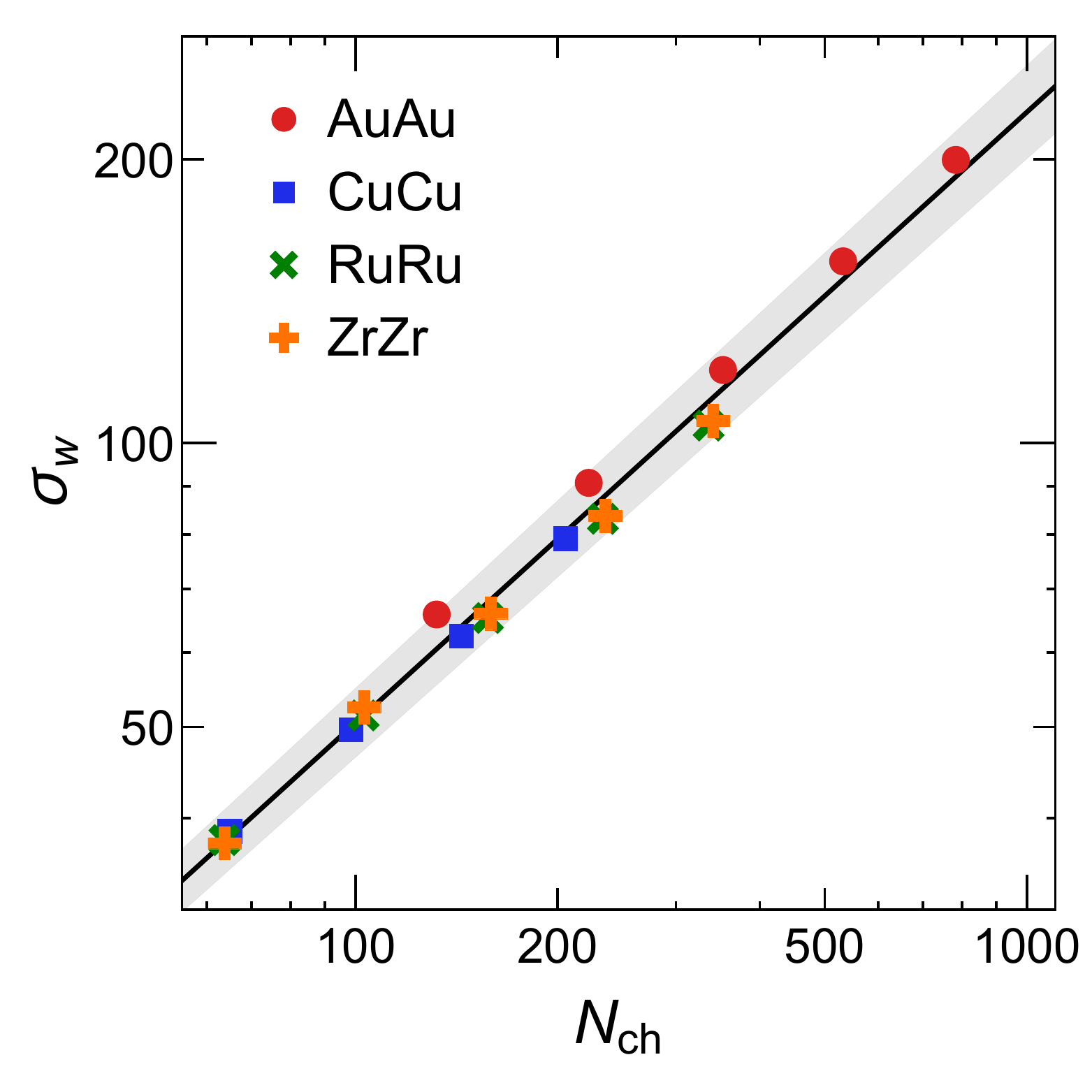}
		\caption{ The  initial topological fluctuations $\mathlarger{\mathlarger{\mathlarger{\sigma}}}_w$  versus the charged hadron multiplicity $N_{ch}$ (on log-log scale) for various centrality and different colliding systems: AuAu (red circle), RuRu (green cross), ZrZr (orange cross) as well as CuCu (blue square). The black line with grey band demonstrates a universal power-law fitting given in Eq.~(\ref{eq_scaling}).  
	}		\vspace{-0.5cm}
		\label{fig_3}
	\end{center}
\end{figure}

Using the  above result for AuAu collisions at  $(20\sim50)\%$  centrality as the benchmark, we next extend the calculations to all centrality classes as well as to other colliding systems at RHIC $200$GeV beam energy, including the CuCu collisions as well as the isobaric RuRu and ZrZr  collisions~\cite{Skokov:2016yrj,Adam:2019fbq}. 
The ratio of   $\mathlarger{\mathlarger{\mathlarger{\sigma}}}_w$ to initial total entropy $S$   is found to increase from central to peripheral collisions and also increases from larger to smaller colliding systems. This behavior implies that when the fireball created in the collisions becomes smaller  (due to either centrality or nucleus size), the $\mathlarger{\mathlarger{\mathlarger{\sigma}}}_w$ decreases less sensitively than the entropy $S$. Physically, the $\mathlarger{\mathlarger{\mathlarger{\sigma}}}_w$ is correlated with initial seeds of local gluon field flux tubes, the number of which  roughly scales with   the binary collision number $N_{binary}$. The $N_{binary}$ is not directly measurable but is closely correlated with   measurable charged hadron multiplicity $N_{ch}$. To gain further insight, we plot in Fig.~\ref{fig_3} the dependence of $\mathlarger{\mathlarger{\mathlarger{\sigma}}}_w$ on $N_{ch}$ with all centrality and all colliding systems from our computations.   Remarkably the results demonstrate a universal scaling behavior,  as is visible from the black straight line on the log-log scale. A fitting analysis reveals the following  scaling relation:    
\begin{eqnarray} \label{eq_scaling}  
\mathlarger{\mathlarger{\mathlarger{\sigma}}}_w &=& \left( 2.56\pm 0.17 \right)  \times  \ N_{ch}^{\,\, (0.648\pm 0.013)}    \,\, .  
\end{eqnarray}
It would be exciting to test this novel finding with future experimental analyses from all these colliding systems.


 {
Finally, we utilize our finding above to analyze the isobar collisions and offer insights into the interpretation of the latest isobar data from STAR Collaboration~\cite{STAR:2021mii}, specifically focusing on the $(20\sim 50)\%$ centrality class. Based on the measured multiplicities and our finding in Eq.~(\ref{eq_scaling}), one obtains $\mathlarger{\mathlarger{\mathlarger{\sigma}}}_w^{Ru}=36.32$ and $\mathlarger{\mathlarger{\mathlarger{\sigma}}}_w^{Zr}=35.35$. Using these inputs for AVFD simulations, one arrives at the following predictions for CME signals: $H^{Ru}=1.41\times 10^{-5}$ and $H^{Zr}=1.17\times 10^{-5}$, with a signal ratio $H^{Ru}/H^{Zr} = 1.2$. For a reasonable estimate of the background correlations into the experimental observable known as the $\Delta \gamma$-correlator, one could utilize an approximate scaling relation $\Delta \gamma^{bkg} \simeq \frac{g v_2}{N_{ch}}$~\cite{Choudhury:2021jwd,An:2021wof,Kharzeev:2022hqz} where the coefficient $g$ is estimated to be $0.41$ by a comparison with experimental data for $\Delta \gamma = 2 H + \Delta \gamma^{bkg}  $ consistently in both isobar systems. One can therefore further obtain the CME signal fraction $f_{CME}$ in the overall $\Delta \gamma$-correlator of both isobar systems, with 
$f_{CME}^{Ru} \approx 0.040$ and $f_{CME}^{Zr} \approx 0.032$, which appear to be consistent with an independent analysis in \cite{Kharzeev:2022hqz}. It however should be noted that a definitive answer for any potential CME signal in isobar data is still lacking, due to a range of uncertainties involved in both theoretical baseline estimates and experimental analyses.
}

\section*{Summary and Discussion}


In summary, we report a newly developed framework that has  implemented a key ingredient, the stochastic dynamics of gauge field topological fluctuations, into the EBE-AVFD simulations for CME transport in these collisions.Such a framework provides the necessary tool to quantify initial topological fluctuations of QCD gluon fields  from any definitive CME signal to be extracted from experimental measurements of the chiral magnetic effect in relativistic heavy ion collisions. By further applying this tool toward a variety of centrality class and colliding systems, a universal scaling relation between the initial gluon field topology fluctuations and the final charged hadron multiplicity has been identified in Eq.(\ref{eq_scaling}), which can  be tested by experimental data. With combined theoretical and experimental efforts in the near future,  one can look forward to opportunities for extracting the relevant  topological transition rate and understanding the   mechanism underlying nontrivial phenomena such as cosmic baryon asymmetry as well as spontaneous chiral symmetry breaking in QCD vacuum.

\vspace{0.1in}
\section*{Acknowledgments}
We thank A. Tang, F. Wang and G. Wang for useful discussions and communications. A.H. is supported by the Fundamental Research Funds for the Central Universities.   S.S. acknowledges support   
by the Natural Sciences and Engineering Research Council of Canada  and by the Fonds de recherche du Qu\'ebec - Nature et technologies (FRQNT) through the Programmede Bourses d'Excellencepour \'Etudiants \'Etrangers (PBEEE) scholarship. S.L. is in part supported by NSFC under Grant No. 12075328, 11735007 and 11675274. The work of X.G. is partly supported by NSFC Grants No. 12035007 and 11905066 as well as by Guangdong Major Project of Basic and Applied Basic Research No. 2020B0301030008 and Science and Technology Program of Guangzhou No. 2019050001.   
The research of J.L. is supported by the NSF Grant No. PHY-2209183.     
 
%
\bibliographystyle{unsrt}
\bibliography{cme.bib} 



%

\end{document}